\documentclass[10pt]{article}

 \usepackage{amsmath,amsthm,amssymb}
 \usepackage{makeidx,epsfig,lscape}
 \usepackage{color,colortbl}
 \usepackage{fancyhdr}
 \usepackage{xcolor,pict2e}

 \definecolor{myaqua}{rgb}{0.0,0.5,0.55}
 \definecolor{lightaqua}{rgb}{0.75,0.95,0.95}

 \usepackage[colorlinks = true,
            linkcolor = myaqua,
            urlcolor  = blue,
            citecolor = myaqua]{hyperref}

\usepackage{caption}
\usepackage{floatrow}

\def\lin#1#2{\textcolor[rgb]{0.6,0.6,0.6}{\vspace*{#1mm} \hrule
   height 3 pt \vspace*{#2mm}}}
\def\bt{\begin{tabular}}
\def\et{\end{tabular}}
\def\and{\mbox{ and }}

\def\1{{\bf 1}}

 \def\sectionn#1{\refstepcounter{section}{\color{myaqua}

 \vskip 6mm

 \noindent\Large\bf\thesection. #1}

 \vskip 3mm}
 \def\subsectionn#1{\refstepcounter{subsection}{\color{myaqua}

 \vskip 5mm

 \noindent\large\bf\thesubsection. #1}

 \vskip 2mm}

\begin{document}


 $\mbox{ }$

 \vskip 12mm

{ 

{\noindent{\huge\bf\color{myaqua}
 Microscale crystalline rare-earth doped resonators \\[2mm] for strain-coupled optomechanics}}
%
\\[6mm]
{\large\bf J.F. Motte$^1$, N. Galland$^1$, J. Debray$^1$, A. Ferrier$^{2,3}$, P. Goldner$^2$, N. Lu{\v c}i\'c$^4$, S.~Zhang$^4$, B. Fang$^4$, Y. Le Coq$^4$, S. Seidelin$^{1,5}$}}
\\[2mm]
{ 
 $^1$Univ. Grenoble Alpes, CNRS, Grenoble INP and Institut N\' eel, 38000 Grenoble, France\\
$^2$ PSL Research University, Chimie ParisTech, CNRS, Institut de Recherche de Chimie Paris, 75005, Paris, France\\
$^3$Sorbonne Universit\'es, UPMC Universit\'e Paris 06, 75005, Paris, France\\
$^4$ LNE-SYRTE, Observatoire de Paris, Universit\' e PSL, CNRS, Sorbonne Universit\' e, Paris, France\\
$^5$Institut Universitaire de France, 103 Boulevard Saint-Michel, 75005 Paris, France\\
Email:
\href{mailto:signe.seidelin@neel.cnrs.fr}{\color{blue}{\underline{\smash{signe.seidelin@neel.cnrs.fr}}}}
 \\[4mm]

\lin{5}{7}

 { 
 {\noindent{\large\bf\color{myaqua} Abstract}{\bf \\[3mm]
 \textup{Rare-earth ion doped crystals for hybrid quantum technologies is an area of growing interest in the solid-state physics community. We have earlier theoretically proposed a hybrid scheme of a mechanical resonator which is fabricated out of a rare-earth doped mono-crystalline structure. The rare-earth ion dopants have absorption energies which are sensitive to crystal strain, and it is thus possible to couple the ions to the bending motion of the crystal cantilever. Here, we present the design and fabrication method based on focused-ion-beam etching techniques which we have successfully employed in order to create such microscale resonators, as well as the design of the environment which will allow to study the quantum behavior of the resonators. }}}
 \\[4mm]
 {\noindent{\large\bf\color{myaqua} Keywords}{\bf \\[3mm]
Rare-earth Doped Crystals; Mechanical Resonators; Optomechanics; Quantum Physics; Strain-coupling; Spectral Hole Burning; Focused-Ion-Beam Etching Techniques.}


\lin{3}{1}

\sectionn{Introduction}

{ \fontfamily{times}\selectfont
 \noindent The study of mechanical resonators coupled to light belongs to the field of optomechanics \cite{Aspelmeyer2012}, which emerged experimentally about a decade ago, when several groups started to investigate the techniques required to actively cool a macroscopic mechanical oscillator down to its quantum ground state. The combination of active and traditional cryogenic cooling techniques was intensively pursued and allowed in 2011 for the very first time to place a mechanical system in its quantum ground state \cite{Chan2011,Teufel2011}. One way to study the resonators in or close to the quantum ground state, is to couple it to a two-level system, and interact with the resonator via this system. One particularly interesting way to achieve this is to use ``strain-coupling'', first demonstrated in 2014 with a semi-conductor nanowire \cite{Yeo2014}. In such a resonator, the vibrations, which can be induced deliberately by means of a piezo actuator or result from the Brownian motion due to a finite temperature (or even from the zero-point energy position fluctuations), generate a mechanical strain, as illustrated in Fig.\ref{strain}. This strain influences the electronic properties of the impurity, as a consequence of the modification of the electronic orbital distributions.  The oscillations of the crystal are therefore mapped onto the energy levels of the impurity, which in turn gives rise to a change in the optical frequency of the photons absorbed and emitted. The corresponding strain mediated coupling strength can be higher, and is potentially more stable, than what can be achieved with any realistic external forces (such as magnetic gradient forces \cite{arcizet2011}). What is particularly appealing about these integrated hybrid mechanical systems is that some of them may approach or even enter the strong coupling regime \cite{Dalibard1992}. In this regime, the hybrid coupling strength exceeds the decoherence rates of both the mechanical resonator and the impurity (which, according to context, is the decoherence rate of either an electronic transition or a transition between spin states).\\
 
 \begin{figure}[h]
\centering
\includegraphics[width=160mm]{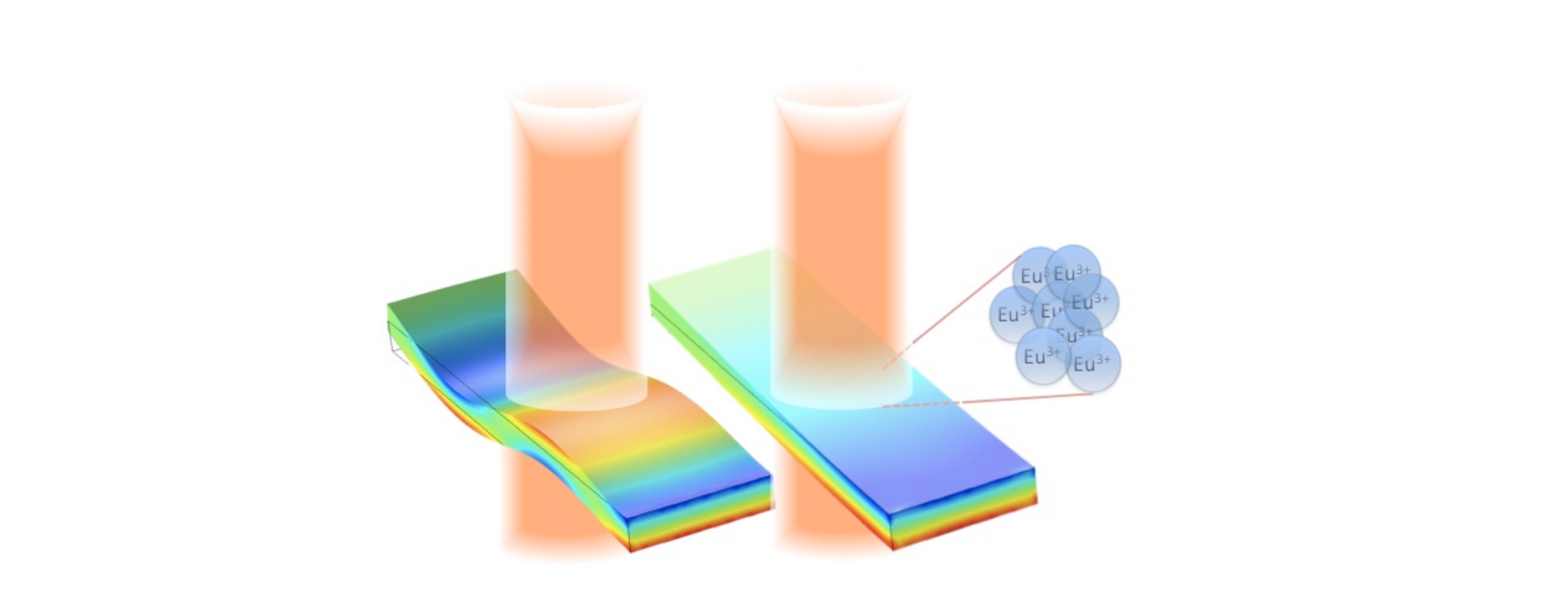}
\caption{\label{strain} When a mechanical resonator vibrates, the strain inside the material is periodically modulated. In this example, the color range from blue (compression) over green (non-strained) to red (extension) of the material is obtained from a COMSOL simulation. The anchoring point (not shown) of the resonators is in foreground of the drawing (toward the reader). To the right, the resonator oscillates in the first excited mode whereas to the left, a higher order excited mode is shown. The energy levels of the europium ions inside the resonator are sensitive to the strain, and these levels can be probed with a laser beam.
}
\end{figure}

This strain-coupling can also be exploited in a rare-earth ion doped crystal resonator, which are particularly interesting due to the exceptional coherence properties of the rare-earth ion dopants. We suggested a theoretical protocol for creating and operating such a strain-coupled system \cite{moelmer2016} and for cooling the resonator down to the near the quantum groundstate \cite{seidelin2019}. In the following, we will discuss how to design the physical system allowing to perform such experiments, with particular attention given to the fabrication of the microscale resonators.

\sectionn{Preliminary theoretical considerations}

In order to develop a resonator design based on Eu$^{3+}$ doped Y$_2$SiO$_5$ for studying strain-coupled optomechanics, several challenges must be faced. First, the atomic structure make it very challenging to observe single ions \cite{casabone2018} due to the narrow absorption linewidth, weak oscillator strength, and the fact that there is often no closed transitions \cite{Equall1994}. We therefore decided to investigate the possibility of using not single ions, but spectral holes imprinted in the structure to which couple the mechanical motion. This is indeed possible, but the fact that an ensemble of ions is involved, adds an inhomogeneous broadening of the lineshape (and not just a linear shift) due to the strain arising from vibrations \cite{moelmer2016}. In other words, the spectral hole broadens under applied strain. In order to overcome this challenge, our protocol is based on a dispersive coupling between the edge of a spectral hole and a detuned laser. More precisely, when the resonator vibrates, the slope of the edge of the spectral hole will oscillate periodically. The coupling can be read-out be observing a phase shift of a laser traversing the resonator. The details of the coupling mechanism are given in Ref. \cite{moelmer2016}, including a protocol for ``functionalizing'' the spectral hole in order to enhance the coupling.

\subsectionn{Parameters obtained from simulations}

One of the practical consequences of coupling to an ensemble of ions is that laser needs to interact with a large number of ions, making a micrometre scale resonator more suitable than a nanoscale one, at least for the initial experiments during which the detection efficiency has not yet been fully optimized. For the first prototypes, based on the simulations, we therefore opted for the following parameters: a single-clamped cantilever with the dimensions 100 $\times$ 10 $\times$ 10 $\mu$m$^3$ interacting with a laser beam traversing the cantilever near its fixed end (anchoring point) for maximum strain. The cantilever which consists of Y$_2$SiO$_5$ (which has a Young Modulus of 135 GP) with an effective mass of 1.1 $\times$ 10$^{-11}$ kg, and of which the first excited mechanical mode vibrates at 890 kHz (value obtained from COMSOL simulations). The cantilever contains a 0.1 \% doping of Eu$^{3+}$ ions, with a $^7$F$_0$ $\rightarrow$ $^5$D$_0$ transition centered at 580 nm and potentially with a linewidth as low as $\Gamma=2\pi \times$122 Hz (at T=1.4 K and 10 mT magnetic field \cite{Equall1994}). For our simulations, we used a power of 1 mW and a hole width of 6 MHz. In this configuration, the static displacement of the tip of the resonator due to the light field amounts to 0.4 pm and the corresponding phase shift of the laser (the carrier) equals 0.2 mrad. This shift is easily observable as, for the 1 mW laser power, the shot noise limited phase resolution is 0.45 microradian within the allowed detection time, before ``hole-overburning'' becomes non-negligible (approximately 16 ms for the 122 Hz linewidth). For comparison, a direct reflection of a 1 mW laser on the resonator (creating a force on the resonator due to the momentum delivered by the photons) would give rise to a much smaller static displacement (20 fm), justifying the efficiency of this dispersive approach. \\

Calculations also showed that the amplitude of the Brownian motion at 3 K is 0.2 pm, and the spectral sidebands of the detection laser contain an integrated phase of 0.11 mrad due to this thermal excitation. For an integration time equal to the inverse of the thermal linewidth (25 microseconds), the shot-noise limited phase-resolution is 14 microradian. The thermally excited sidebands are therefore readily observed, even within such short integration time. Moreover, by increasing the integration time up to the maximum before over-burning, it is possible to observe and measure accurately the detailed shape and size of the sidebands. Zero-point motion of the resonator, averaged over the measurement, induces a small excess integrated phase of 0.4 microradian in the sidebands. As this value is close to the phase resolution achieved within the maximum integration time before hole-overburning, the shot noise limited resolution is therefore sufficient to observe the effect of the zero-point motion of the mechanical resonator. By using either dilution fridge or an active laser cooling mechanism (or a combination of these), the temperature can be lowered near a point where the thermal excitation does not hide the effect of zero point fluctuations. For example, at 30 mK, the zero-point fluctuation induces approximately $10^{-3}$ relative excess integrated phase over the effect of Brownian fluctuations alone. Such a deviation seems measurable, provided sufficient knowledge of the relevant parameters (temperature, Q factor,...). Several measurements at different temperatures may also be used to estimate the various relevant parameters with the necessary accuracy. Note that the resolution can be further increased by repeating the full hole imprinting and measurement sequence several times, or use optical repumpers to preserve the spectral hole \cite{cook2015}. As the current setup is based on a 3 Kelvin cryostat, the zero-point motion measurement are not immediately feasible, but we have here included the numbers as this is one of the next major modifications.\\

A potentially perturbing effect arises due to fluctuations of the laser power. As the static displacement corresponding to 1 mW of laser power is approximately 0.4 pm, this laser power must be stable to within $10^{-4}$ to ensure a perturbation much smaller than the zero point fluctuations (approx. 1 fm), a power stability requirement well within reach of standard stabilization techniques. In addition, the laser frequency must also be carefully controlled. The zero-point motion induces a frequency shift of the ions closest to the edge of the resonator of approximately 37 Hz. The probe laser must therefore have a frequency stability substantially better than that, which is well within reach of nowadays commercially available ultra-high finesse Fabry-Perot cavity stabilized lasers (which typically have sub-Hz frequency stability). Note that probing the Brownian motion alone at 3 K exhibits a much less stringent frequency stability requirement at the sub-kHz level, and this will be one of the first measurements we plan to perform using the resonators. In order to anticipate the zero-point motion measurement, a new commercial ultrastable cavity (Stable Lasers) has recently been integrated in the setup, see section \ref{sec:optics} below.

\sectionn{Microfabrication of resonators}

{ \fontfamily{times}\selectfont
 \noindent

In this section, we will discuss the fabrication of the resonators, of which the design is based on the above considerations. The bulk crystals are grown at the Paris Institute for Chemical Research (Institut de Recherche de Chimie Paris), and the microfabrication of the resonators in a cleanroom facility in Grenoble (Institut NEEL and CEA).

\subsectionn{Growing of the bulk crystals}

A single Y$_2$SiO$_5$ crystal of  0.1 percent at. Eu$^{3+}$ (Eu:YSO) was grown using a homemade RF induction Czochralski apparatus using Ir crucible to contain the melt and ZrO$_2$ and Al$_2$O$_3$ as refractory materials. The growth of the single crystal was performed under flowed nitrogen introduced as a protective atmosphere. During the growth the shape of the crystal is controlled with a feedback loop on the RF generator power by monitoring the weight of the crystal.   
The crystal was pulled at 0.5 mm/h from the melt in the [010] direction, which minimizes the defects due to the anisotropy of the expansion coefficients. The resulting one-inch diameter single crystal is transparent, colorless with a regular shape. The crystal is then oriented by the Laue method with an accuracy of $\pm$ 2 degrees. Finally, the samples were polished with a Logitech PM5. More details on the fabrication of the bulk crystals can be found in Ref. \cite{ferrier2016}.

\subsectionn{Fabrication of microresonators}

The bulk crystal samples have dimensions of the order of millimeters. In order to reach quantum regimes, resonators with micro- and nanoscale dimensions are required. These designs can be realized by subsequently using Focused Ion Beam (FIB) techniques to shape the crystal. Etching a Y$_2$SiO$_5$ crystal using FIB techniques is challenging, as the crystal is an oxide and thus not conducting, leading to rather low etching rates. The literature on the subject is sparse, and a major effort has been required in order to calibrate the procedure and achieve a good etching efficiency. We will in the following outline the essentiel steps and parameters.\\

We use a FIB-SEM system ({\it Zeiss NVision}) with a gallium liquid ion source. Due to the absence of conductivity of the Y$_2$SiO$_5$ it is necessary to first deposit a thin (100-150 nm) metal layer (aluminium) on the crystal prior to etching the structures. This layer prevent the charges from building up, which otherwise would cause the ion-beam to be deviated from the desired target. Due to the hardness of the crystal, a rather high voltage is required to etch the structures, and we operate the gallium ion source at 30 kV (maximum voltage). At this voltage amorphizing of the material can happen, but due to the nature and hardness of the crystal, this turns out not to be an issue for our samples. The current used is approximately 6 nA. We obtain an etching rate of the order of 1 $\mu$m$^3$/nA/s.\\

Once the calibration completed, we turn to the fabrication of the resonators. The major challenge is to etch deep enough into the material in order to detach the resonator from below (we wish to obtain a free-standing resonator, only attached to the crystal at one anchoring point). This requires a large angle of the incoming gallium ion beam, and the ability to remove a large amount of material. As the etching rate is slow, and the extent of orientation of the sample relative to the ion beam is limited, this way of proceeding turns out not to be ideal. \\

\begin{figure}[h]
\centering
\includegraphics[width=160mm]{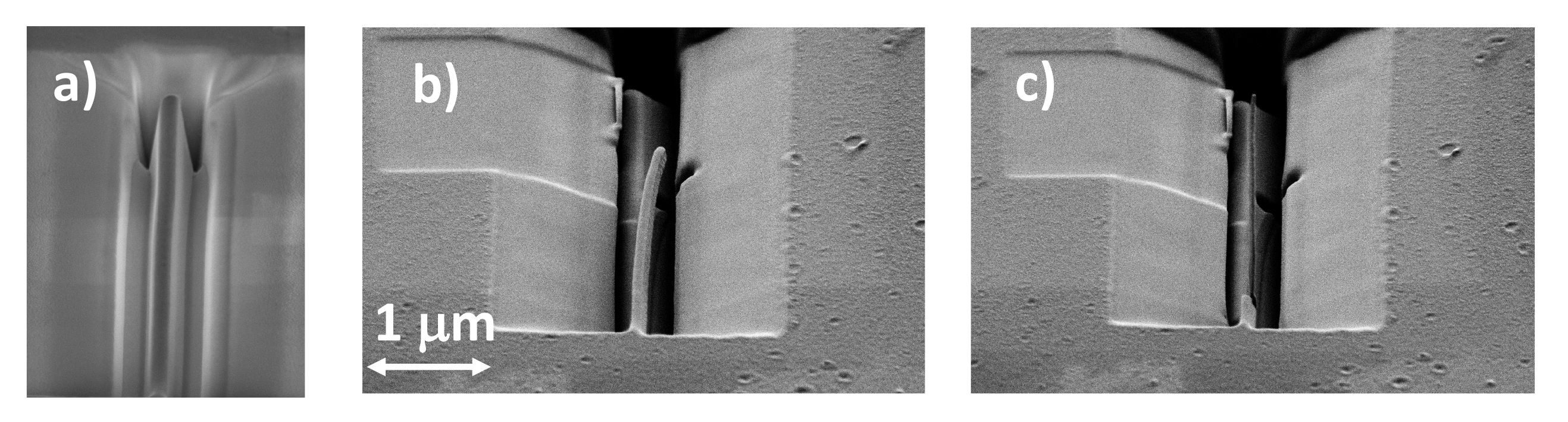}
\caption{\label{trials} In a) we show the resonator etched on the sides, but not liberated from below. In b) we have increased the angle of the gallium ion beam, which allows to etch below the structure. In c) the resonator breaks during an attempt to liberate it from below. In all 3 images, the voltage used for SEM imaging is 2.00 kV.
}
\end{figure}

Figure \ref{trials} a shows an example of a resonator of which the contour lines have been easily etched, but illustrates the difficulty of liberating the resonator from below. In addition to the difficulty of removing material and obtaining the right angle for the gallium ion beam, etching deep trenches are also problematic due to the fact that the metal layer (deposited for removal of accumulated charges) becomes spatially further and further separated from the part of the structure to be etched. By modifying the angle of the gallium ion beam, we managed to liberate most of the resonator from the substrate (see fig.~\ref{trials} b), but the exerted force of the gallium beam and built-up residual charges distort the shape of the resonator, and when proceeding to liberate the resonator near its anchoring point, it breaks off, as illustrated in fig.~\ref{trials} c.\\

A solution which circumvents the need for etching below the resonator, consists in creating the resonator on the edge of a corner the crystal, and this is the technique that has turned out to be the most successful, as it requires the minimal etching of material. Moreover, by turning the crystal 90 degrees the ion-beam can be perpendicular to the crystal at all times. This has been carried out on the crystal shown in fig.~\ref{resonator} a. We have etched a part of the bulk crystal at an angle of 45 degrees below the resonator. By coating this surface with aluminum, we have realized a mirror which allows a laser beam to be reflected onto it and pass through the resonator alone, without interacting with the bulk material of the crystal. We have chosen to position the mirror below the part of the resonator which is nearest the anchoring point in order to maximize the material strain, for an optimum coupling.\\

An alternative option consists in polishing the edge of the crystal down to obtain a very thin layer prior to applying the FIB. As shown in the example in  fig.~\ref{resonator} b, the crystal has been polished to an angle of 3.2 ($\pm$ 0.1) degrees (measured with a DEKTAK profilometer and a calibrated microscope). This allows one to create the resonator by applying the FIB perpendicular to the surface of the crystal along a single direction. This has the advantage of allowing the laser beam to pass directly through the resonator in a straight line, without having to integrate a mirror to deviate the beam to avoid the bulk part of the crystal. Moreover, it also represents the advantage of not having to used the FIB on the top and bottom surface of the resonator, and thus conserve the initially polished surfaces. In that way, we also avoid surface effects of the FIB due a non-desired implementation of the gallium ions, although this has been shown not to represent major degradation of the coherence properties of other species of rare-earth ions in the YSO matrix \cite{zhong2015}.  If however, the triangular shape of the resonator turns out to be a limiting factor (as it increases the resonator's stiffness) it is also possible to reshape it into a rectangular form using the FIB, by removing a much smaller amount of material than when starting out from the bulk crystal.\\

In both resonators, the crystal has been oriented such that the laser beam is parallel to the crystal b-axis (corresponding the the [010] growth axis of the crystal) which optimizes the absorption of the Eu$^{3+}$ ions for the crystal site 1 (there exists 2 non-equivalent crystal site for the ions). The polarization of the laser beam is roughly along the D1 direction, but is fine-tuned manually by maximizing the absorption of the europium ions~\cite{Traum2013}.

\begin{figure}[h]
\centering
\includegraphics[width=160mm]{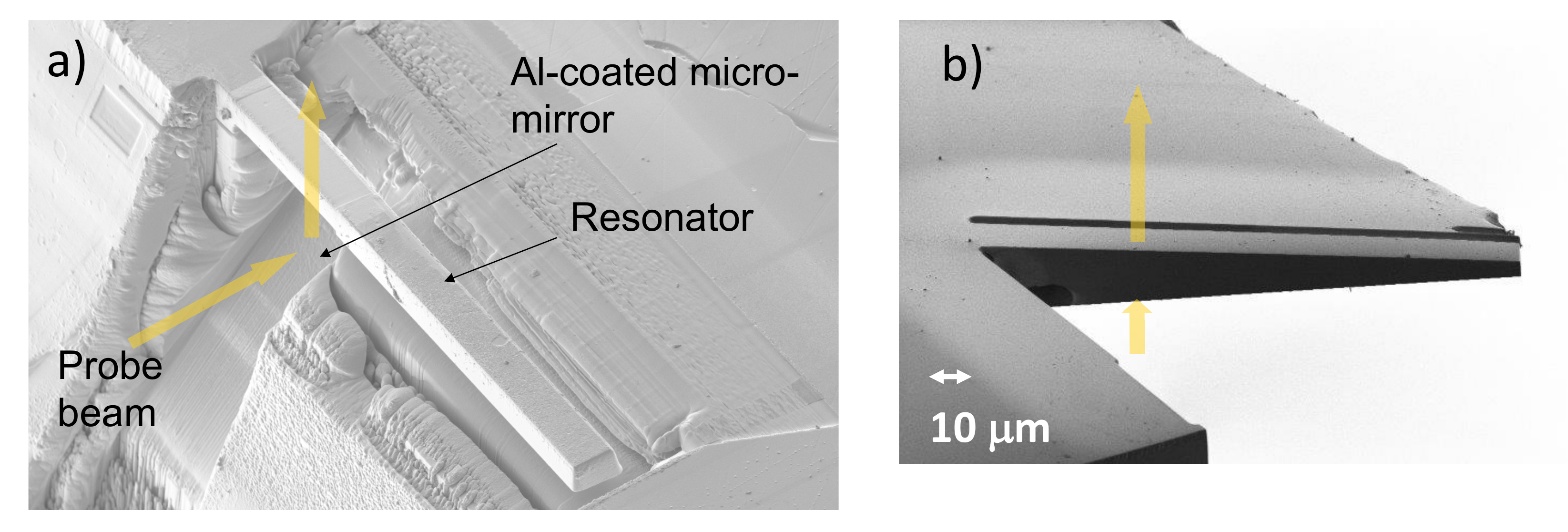}
\caption{\label{resonator} In a) we show the SEM image of the resonator etched on the corner of the crystal. A mirror at an angle of 45 degrees is integrated in order to make the laser beam traverse only the resonator and thus avoiding the bulk part of the crystal.  In b) the edge of the crystal have been polished so thin that the resonator is formed by a single vertical cut with the FIB. The scale indicated in b) applies to both images. 
}
\end{figure}

\sectionn{Optics environment}
\label{sec:optics}

The optical system which will be used to probe the europium ions inside the resonators is based on an extended cavity diode laser at 1160~nm (Toptica DLPro), delivering 65~mW. It is coupled and frequency doubled in PPLN waveguides with free space outputs to reach 580~nm, corresponding to the wavelength of absorption of the $^7F_0 - ^5D_0$ transition in $\rm Eu^{3+}{:}Y_2SiO_5$. The laser at 1160~nm  is frequency locked by the Pound-Drever-Hall method to a commercial reference cavity (Stable Laser Systems). Both the diode laser current and a piezo actuator acting on the external cavity length are used for feedback, with a bandwidth around 500~kHz. After the optical frequency doublers, approximately 10 mW of 580~nm light is obtained. Details on the laser system can be found in Ref.~\cite{Gobron2017}.\\

\subsectionn{Optics mount for studying the resonator}

We present our setup for focussing a laser beam onto the micrometric resonator, taking into account the constraints of the cryogenic environment. Our cryostat (OptiDry200 from MyCryoFirm) has a fairly large (approximately $35$~cm in external diameter) vacuum chamber, with an hermetically sealed inner vacuum chamber to allow for the possibility of cold buffer gas at around $4$~K. Together with a thermal shield at $50$~K, there are in total three view ports (fused silica of varying thickness between $4$ to $10$~mm) in series between the outer wall and the sample holder where the resonator will be installed. In order to minimize aberrations, it is preferable to propagate a collimated laser beam through these windows and focus down to a spot of $<5~\mu$m (RMS spot size) near the resonator, the width of which is $10~\mu$m. The beam after passing through the resonator will thus be highly diverging, and will be collimated again in the cryostat before exiting and reaching the detection optics. In addition, an afocal telescope of unit magnification (Thorlabs LA1131A, $f = 50$~mm) is placed before the cryostat to allow for potential fine adjustment of the collimation.\\

\begin{figure}[h]
\centering
\includegraphics[width=160mm]{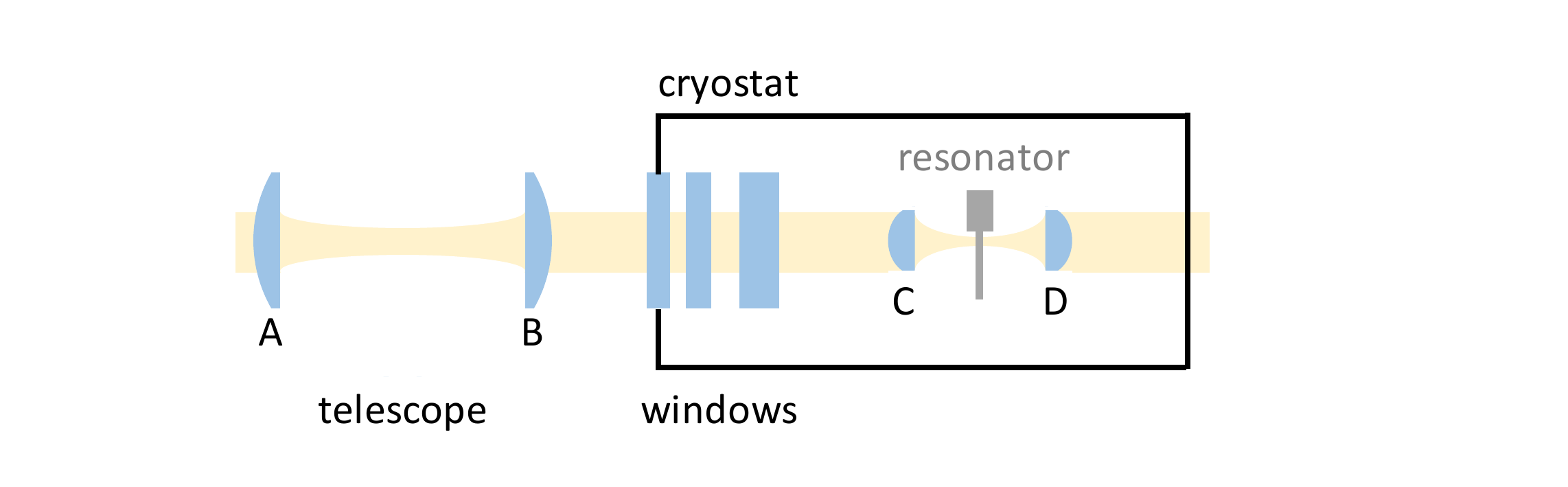}
\caption{ \label{fig:optsetup} A schematic of the optical set up: an afocal telescope (lenses A and B) of unit magnification, the 3 windows of the cryostat, the focusing lens (lens C), the resonator, and the collimating lens (lens D). Lens B is allowed to move in the following simulations to compensate for any possible defocus. Note that the detailed structure of the cryostat, the exit view ports, and the detection optics are not represented in this figure.}
\end{figure}
 
We simulate using OSLO (ray tracing software from Lambda Research Corporation) the RMS spot size of a Gaussian beam ($1/e^2$ radius of $2.75$~mm) and its ensquared energy within a square of $10~\mu$m $\times$ $10~\mu$m at the position of the resonator. We also consider a defocus of $\pm 5$ $\mu$m to account for the thickness of the resonator). In an ideal situation, i.e. the resonator ($10~\mu$m $\times$ $10~\mu$m in cross section) is placed exactly at the focal point of lenses C, Thorlabs asphere C280TMD-A, $f=18.4$~mm, the RMS spot size of the laser beam remains below $2~\mu$m, to be compared with a diffraction limited Airy radius $2.9~\mu$m. The total ensquared energy is more than $93\%$ for distance of $\pm 5~\mu$m from the focal point. In other words, there will be little stray light elsewhere that may contribute to a background noise. \\

A first optomechanical mount has been realised for the resonator (which is depicted in Fig.~\ref{resonator} a), containing the lenses C and D. The geometry of this resonator requires an additional mirror (Thorlabs prism MRA05-E02) at $45^\circ$ to deflect the outgoing laser beam in order to use the view ports in a straight through configuration. For this purpose, the optical axes of the focusing (C) and collimating (D) lenses are displaced. Besides the focal position of the lenses, there are no adjustment available once the resonator is fixed.  We modify our simulation to investigate the effect of 1) a possible defocus (from longitudinal positioning and thermal contraction of the mount during the cooling cycle), and 2) a transverse displacement (mainly due to finite positioning accuracy). \\
 
Given that the integrated fractional thermal contraction of copper from room temperature down to $4$~K is about $3\times 10^{-3}$, a lens-resonator distance (longitudinal) on the order of $20$~mm would vary about 60 microns. We verify that the afocal telescope can conveniently accommodate a defocus of $\pm 100~\mu$m on axis. Typically, a defocus of $10~\mu$m requires to adjust the position of lens B by about hundreds of microns, easily achievable with a manual translation stage possessing a $5$~mm course. After compensation, the RMS spot size remains below $2~\mu$m and the ensquared energy is above $90\%$.\\

We also consider a transverse positioning error of $\pm 50~\mu$m, both at the best focus and with a defocus of $\pm 100~\mu$m but compensated by repositioning lens B as described in the previous paragraph. In the worst case, i.e. at the maximal transverse positioning error and defocus $= -100~\mu$m, the RMS spot size remains about $2~\mu$m and the ensquared energy is above $80\%$. \\

The planned optical setup (see Fig.~\ref{fig:optsetup}) is therefore simple, robust and promising for our future experiments, as it can focus most of the light down to a size compatible with the microresonator and is able to accommodate $\pm 100~\mu$m of longitudinal position error and $\pm 50~\mu$m of transverse position error. It will require a pre-alignment at room temperature and pressure prior to its installation in the cryostat. We are confident that the thermal effects of the cooling cycle should not prevent us from obtaining the desired signal from the resonator.\\

{ \fontfamily{times}\selectfont
 \noindent


\sectionn{Conclusion and outlook}

{ \fontfamily{times}\selectfont
 \noindent
 
 In this article, we have motivated why rare-earth doped materials in shapes of resonators might hold promise for quantum optomechanics experiments, and we have provided a description, including the most essential physical parameters, allowing one to successfully fabricate such resonators. We have also briefly described the experimental environment including optics and cryogenic setup which meet the requirements for studying the resonators near the quantum regime. We are currently able to observe the desired spectral features in bulk crystals, with a signal to noise level sufficiently high to potentially be able to observe similar signals in the microresonators, which constitutes our next step. Once this achieved, by exciting the resonator with a piezo drive, it should be straightforward to observe the strain-coupling between resonator and europium ions in the classical regime, before proceeding to apply the cooling protocol described in Ref.~\cite{seidelin2019} allowing to reach the quantum ground state.

 \fancyfoot{}
 \fancyfoot[C]{\leavevmode
 \put(0,0){\color{lightaqua}\circle*{34}}
 \put(0,0){\color{myaqua}\circle{34}}
 \put(-5,-3){\color{myaqua}\thepage}}

 {\color{myaqua}

 \vskip 6mm

 \noindent\Large\bf Acknowledgments}

 \vskip 3mm

{ \fontfamily{times}\selectfont
 \noindent
YLC acknowledges support from the Ville de Paris Emergence Program and from the LABEX Cluster of Excellence FIRST-TF (ANR-10-LABX-48-01), within the Program ``Investissements d'Avenir'' operated by the French National Research Agency (ANR). The project has also received funding from the European Union's Horizon 2020 research and innovation program under grant agreement No 712721 (NanOQTech).

 {\color{myaqua}

}}

\end{document}